\documentclass[12pt]{article}
\usepackage{hyperref}

\def\NON{\nonumber\\}
\def\NXT{\\}

\def\a{\alpha}

\def\c{\chi}

\def\e{\epsilon}                % Also, \varepsilon
\def\g{\gamma}
\def\h{\eta}

\def\j{\psi}

\def\m{\mu}
\def\n{\nu}

\def\p{\pi}                     % Also, \varpi
\def\t{\tau}

\def\D{\Delta}

\def\G{\Gamma}

\def\O{\Omega}

\def\ca{{\cal A}}
\def\cb{{\cal B}}

\def\cd{{\cal D}}

\def\cg{{\cal G}}
\def\cm{{\cal M}}
\def\cn{{\cal N}}

\def\cq{{\cal Q}}

\def\ct{{\cal T}}

\def\cw{{\cal W}}

\def\cbo{{\,\raise-.15ex\Sc [\,}}                       % curly "
\def\svev#1{\left\langle #1\right\rangle}       % variable < >
\def\ddt#1{{\buildrel {\hbox{\LARGE .\kern-2pt.}} \over {#1}}}% double dot-over
\def\sstyle{\scriptstyle}

\def\ie{\mbox{\it i.e.} }
\def\eg{\mbox{\it e.g.} }

\def\frac#1#2{ {\sstyle {#1\over #2} } }

\usepackage{amsmath}

\thicklines                         % thick straight lines for pictures (LaTeX)
\def\bj{\overline\psi}
\def\bc{\overline\chi}
\def\bh{\overline\eta}
\def\bs{\overline{S}}
\def\tG{\tilde\G}
\def\tx{\tilde{x}}
\def\ty{\tilde{y}}
\def\tp{\tilde{p}}
\def\Dst{D}
\def\abss#1{\parallel\! #1 \!\parallel}
\def\textit#1{{\it \!\!\! #1 \!\!}}
\def\kn{k^{(n)}}

\begin{document}
\hyphenation{fer-mio-nic per-tur-ba-tive pa-ra-me-tri-za-tion
pa-ra-me-tri-zed a-nom-al-ous}

\begin{center}
\vspace{10mm}
{\large\bf Locality of the fourth root of the Staggered-fermion
\\[2mm]
determinant: renormalization-group approach}
\\[12mm]
Yigal Shamir
\\[8mm]
{\small\it School of Physics and Astronomy\\
Raymond and Beverly Sackler Faculty of Exact Sciences\\
Tel-Aviv University, Ramat~Aviv,~69978~ISRAEL}\\
{\tt shamir@post.tau.ac.il}
\\[10mm]
{ABSTRACT}
\\[2mm]
\end{center}

\begin{quotation}
Consistency of present-day lattice QCD simulations with dynamical (``sea'')
staggered fermions requires that the determinant of the
staggered-fermion Dirac operator, ${\rm det}(\Dst)$, be equal to
${\rm det}^4(D_{\rm rg})\, {\rm det}(\ct)$
where $D_{\rm rg}$ is a local one-flavor lattice Dirac operator,
and $\ct$ is a local operator containing
only excitations with masses of the order of the cutoff.
Using renormalization-group (RG) block transformations I show that,
in the limit of infinitely many RG steps,
the required decomposition exists for the free staggered operator
in the ``flavor representation.'' The resulting one-flavor Dirac operator
$D_{\rm rg}$ satisfies the Ginsparg-Wilson relation in the massless case.
I discuss the generalization of this result to the interacting theory.
\end{quotation}

%%%%%%%%%%%%
\newpage
\noindent {\large\bf 1.~Introduction}
%\secteq{1}
\vspace{3ex}

Lattice QCD simulations with dynamical (``sea'') staggered fermions \cite{ks}
are providing predictions for hadronic observables with unprecedented
accuracy \cite{milc}. In these numerical calculations, it is crucial that
\textit{all} sources of error be under systematic control.
This raises the question of the validity of the ``fourth root trick''
used in these simulations.

Let me briefly explain the problem (more details may be found \eg
in ref.~\cite{rev}). In four dimensions, the staggered Dirac operator $D$
is a one-component lattice operator which, in the free-field case,
has sixteen poles in the Brillouin zone.
These poles combine into four Dirac fermions (with a total of sixteen
degrees of freedom) in the continuum limit.
To account for three dynamical quarks -- up, down and strange --
the Boltzmann weight used for generating the dynamical configurations
involves the factor\footnote{
  In practice, $m_u=m_d$ in the simulations. Note also that the simulations
  correspond to a ``hybrid'' and/or partially-quenched theory,
  since the ``sea'' and the ``valence'' quarks may differ in their masses,
  as well as in the details of the discretization used for each.
}
\begin{equation}
  {\rm det}^{1/4}(\Dst(m_u))\; {\rm det}^{1/4}(\Dst(m_d))\;
  {\rm det}^{1/4}(\Dst(m_s)) \,.
\label{fourth}
\end{equation}
Taking the fourth root of each staggered-fermion determinant
ensures that the lattice theory
describes three (and not twelve) quarks in the continuum limit.
While the ``fourth root trick'' is necessary in practice in order to reach
the desired continuum theory, it is not obvious that this trick is consistent.
The question is whether the gauge-field configurations generated with this
Boltzmann weight correspond to a {\it local} lattice theory.
If one could show that ${\rm det}(\Dst)$ is equal to the fourth
power of the determinant of some
local one-flavor lattice Dirac operator,\footnote{
  In this paper, a local operator means an operator whose kernel
  decays exponentially with the separation $|x-y|$,
  with a decay rate which is $O(1)$ in lattice units.
  A similar notion of locality applies to the effective action $S_{\rm eff}(U)$
  discussed below Eq.~(\ref{decomp}).
}
this would provide a positive answer.

In fact, a far \textit{weaker} condition is sufficient to guarantee locality,
and, hence, consistency.\footnote{This observation was recently made
  in ref.~\cite{adams}.
}
Suppose one can show that
\begin{equation}
  {\rm det}(\Dst) = {\rm det}^4(D_{\rm rg})\, {\rm det}(\ct) \,.
\label{decomp}
\end{equation}
Here $D_{\rm rg}$ is a local one-flavor lattice Dirac operator,
and $\ct$ is a local operator containing
only excitations with masses of the order of the cutoff.
We may now write ${\rm det}(\ct) = \exp(-S_{\rm eff}(U))$.
%  where $S_{\rm eff}(U) = \sum_l S^{(l)}_{\rm eff}(U)$,
%  and $S^{(l)}_{\rm eff}(U)$ is a sum over Wilson loops of length $l$.
Since $\ct$ contains only excitations with cutoff masses, we expect
that the effective action $S_{\rm eff}(U)$ is local;
trivially, the same is true for $(1/4)S_{\rm eff}(U)$. By eq.~(\ref{decomp}),
the fourth-root trick then amounts to using dynamical fermions
with the local Dirac operator $D_{\rm rg}$, together with the modification
of the gauge-field action by $(1/4)S_{\rm eff}(U)$.\footnote{
  Locality of ${\rm det}^{1/4}(\ct)$ in the free theory is addressed
  in Sect.~3.
}

The natural framework to realize relation~(\ref{decomp}) is through
RG block transformations. After introducing the relevant concepts
in Sect.~2, the free staggered-fermion operator is dealt with in Sect.~3,
which constitutes the main part of this work.
Using the ``flavor representation'' of staggered fermions \cite{fr}
it is shown that the decomposition~(\ref{decomp})
is realized in the limit of infinitely many RG blocking steps.
Central to this discussion is a theorem on the locality of RG-blocked
Wilson fermions proved in ref.~\cite{rgb}; only a trivial amendment is needed
in order to generalize the theorem to the flavor representation
of staggered fermions.  The limiting operator is constructed explicitly.
(For Wilson fermions, see ref.~\cite{w}. See also ref.~\cite{bw} for an
RG treatment of staggered fermions within the one-component formalism.)
Like any fixed-point operator \cite{h},
the limiting operator satisfies the Ginsparg-Wilson (GW) relation \cite{gw,l}.
Sect.~4 contains a discussion of some of the issues
(both theoretical and practical) that arise in the interacting theory.

%%%%%%%%%%%%
\vspace{5ex}
\noindent {\large\bf 2.~Renormalization-group transformations}
%\secteq{1}
\vspace{3ex}

We first consider a general setup, following refs.~\cite{bkw,rgb}.
Starting from a bilinear fermion action with Dirac operator $D_0$,
an RG blocking is introduced via the identities
\begin{subequations}
\label{RG}
\begin{eqnarray}
  Z &=& \int d\j d\bj\; \exp(-\bj D_0 \j)
\label{RGa}
\NXT
  &=& \int d\j d\bj d\c d\bc\; \exp\left(
  -\bj D_0 \j
  - \a(\bc - \bj Q^\dagger)(\c - Q \j) \right)
\label{RGb}
\NXT
  &=& \int d\j d\bj d\c d\bc d\h d\bh\; \exp\left(
  -\bj D_0 \j + \a^{-1}\, \bh \h
  + (\bc - \bj Q^\dagger)\h + \bh(\c - Q \j) \right)
\label{RGc}
\NXT
  &=& {\rm det}^{-1}(G_1) \int d\c d\bc\; \exp(-\bc D_1 \c) \,.
\label{RGd}
\end{eqnarray}
\end{subequations}
Here $\j,\bj$ live on the ``fine'' lattice with spacing $a_0$, whereas $\c,\bc$
(and the auxiliary field $\h,\bh$) live on the ``coarse'' lattice
with spacing $a_1$ equal to an integer multiple of $a_0$.
In this paper we set $a_1=2a_0$.
The RG blocking kernel $Q$ is a rectangular matrix satisfying
\begin{equation}
  QQ^\dagger = c I \,,
\label{Q}
\end{equation}
where $I$ is the identity matrix on the coarse lattice.
%% [Explicitly, this relation reads $\sum_{z'} Q_{xz'} Q^*_{z'y} = c\, \d_{xy}$
%% where $x$ and $y$ ($z'$) are coordinates on the coarse (fine) lattice.]
Explicitly,
\begin{subequations}
\label{DG1}
\begin{eqnarray}
  G_1 &=& \G_1 = (D_0 + \a Q^\dagger Q)^{-1} \,,
\label{G1}
\NXT
  D_1 &=& \a - \a^2 Q G_1 Q^\dagger \,,
\label{D1}
\NXT
  D_1^{-1} &=& \a^{-1} + Q D_0^{-1} Q^\dagger \,.
\label{D1inv}
\end{eqnarray}
\end{subequations}
Eqs.~(\ref{G1}) and~(\ref{D1}) are obtained by integrating over $\j,\bj$
in Eq.~(\ref{RGb}), and Eq.~(\ref{D1inv}) by first integrating over  $\j,\bj$
and then over $\h,\bh$ in Eq.~(\ref{RGc}).
Iterating the blocking transformation we have
\begin{subequations}
\label{DGj}
\begin{eqnarray}
  \G_j &=& (D_{j-1} + \a Q^{(j)\dagger} Q^{(j)})^{-1}  \,,
\label{Gj}
\NXT
  D_j &=& \a - \a^2 Q^{(j)} \G_j Q^{(j)\dagger} \,,
\label{Dj}
\NXT
  D_j^{-1} &=& \a^{-1} + Q^{(j)} D_{j-1}^{-1} Q^{(j)\dagger} \,,
\label{Djinv}
\end{eqnarray}
\end{subequations}
where $Q^{(j)}$ denotes the blocking kernel from the $(j-1)$-th lattice
(with spacing $2^{j-1}a_0$) to the $j$-th lattice (with spacing $2^{j}a_0$).
We may also go directly from the finest to the coarsest lattice:
\begin{subequations}
\label{DGn}
\begin{eqnarray}
  G_n &=& (D_0 + \a_n Q_n^\dagger Q_n)^{-1}  \,,
\label{Gn}
\NXT
  D_n &=& \a_n - \a_n^2 Q_n G_n Q_n^\dagger \,,\qquad
  \a_n = \a \frac{1-c}{1-c^n} \,,
\label{Dn}
\NXT
  D_n^{-1} &=& \a_n^{-1} + Q_n D_0^{-1} Q_n^\dagger \,,
\label{Dninv}
\end{eqnarray}
\end{subequations}
where $Q_n=Q^{(n)} Q^{(n-1)}\cdots Q^{(1)}$.
Eq.~(\ref{Dninv}) follows by iterating Eq.~(\ref{Djinv}) while using Eq.~(\ref{Q}).
The other equations follow by noting that the product of
$n$ blocking transformations can also be represented as a single ``big''
blocking transformation as in Eq.~(\ref{RG}), provided we let $\c,\bc$ live
on the coarsest lattice, and we make the replacements
$\a\to \a_n$, $Q\to Q_n$. Hence each relation in Eq.~(\ref{DGn}) must match
the corresponding one in Eq.~(\ref{DG1}).
%\begin{equation}
%  D_n^{-1} = \a_n^{-1}
%  + Q_n Q_{n-1} \cdots Q_1\, D_0^{-1}\,
%    Q_1^\dagger  Q_2^\dagger \cdots Q_n^\dagger \,, \qquad
%  \a_n = \a \frac{1-c}{1-c^n} \,,
%\label{Dnn}
%\end{equation}
%%%[In the special case $c=1$, one has $\a_n=\a/n$.]

%%%%%%%%%%%%
%\newpage
\vspace{5ex}
\noindent {\large\bf 3.~Free staggered fermions}
%\secteq{3}
\vspace{3ex}

We now turn to free staggered fermions. In the flavor representation \cite{fr}
the staggered Dirac operator has four-component spin and flavor indices,
and is given explicitly by
\begin{equation}
  D_0 = a^{-1} \sum_\m \big((\g_\m \otimes I) \nabla_\m
  + (\g_5 \otimes \t_5 \t_\m) \D_\m \big) + m \,,
\label{Dst}
\end{equation}
where $a$ is the lattice spacing,
$\nabla_\m f(x) = (f(x+\hat\m) -f(x-\hat\m))/2$,
and $\D_\m f(x) = (2f(x) -f(x+\hat\m) -f(x-\hat\m))/2$.
The usual Dirac matrices are $\g_\m$, while the $\t_\m$ constitute another set
of Dirac matrices acting on the flavor index.
(Taken together, $(\g_\m \otimes I)$ and $(\g_5 \otimes \t_\m)$ form a
representation of the eight-dimensional Dirac algebra.)
For $m=0$, $D_0$ is anti-hermitian.
Going to momentum space one has
$D_0^{-1} = \O^{-1} D_0^\dagger$ where
\begin{eqnarray}
  D_0 &=& a^{-1} \sum_\m \big((\g_\m \otimes I) i\sin(p_\m a)
  + (\g_5 \otimes \t_5 \t_\m) (1-\cos(p_\m a) \big) + m \,,
\NON
%  D_0^{-1} &=& \O^{-1} D_0^\dagger \,,
%%  D_0^{-1}(m;a) &=& \O^{-1}\Big(
%%    -a^{-1} \sum_\m \left((\g_\m \otimes I) i\sin(p_\m a)
%%    + (\g_5 \otimes \t_5 \t_\m) (1-\cos(p_\m a)) \right) + 2m \Big)
%\NXT
  \O &=& D_0^\dagger D_0
  = a^{-2} \sum_\m \big( \sin^2(p_\m a) + (1-\cos(p_\m a))^2 \big) + m^2\,.
\label{Dp}
\end{eqnarray}

We will apply $n$ block transformations to the Dirac operator~(\ref{Dst}).
We set $m=0$, hence we may write
\begin{equation}
  D_0^{-1} = D_0^{-1}(p;a) = -\sum_\m \big( i(\g_\m \otimes I) \ca_\m^0(p;a)
  + (\g_5 \otimes \t_5 \t_\m) \cb_\m^0(p;a) \big) \,.
\label{D0inv}
\end{equation}
We will hold fixed the lattice spacing obtained in the
$n$-th step. We thus set $2^n a_0 = 1$, or $a_0 = 2^{-n}$.
The blocking kernel $Q^{(j)}$ is defined as follows.
We label the sites of the $(j-1)$-th lattice by four integers
$l=(l_1,l_2,l_3,l_4)$, $l_\m \in Z$.
A site $l'=(l'_1,l'_2,l'_3,l'_4)$ of the $j$-th lattice
is identified with the site $2l'=(2l_1,2l_2,2l_3,2l_4)$
on the $(j-1)$-th lattice. The blocking transformation assigns to a field
variable on the $j$-th lattice its arithmetic mean over a $2^4$ hypercube
on the $(j-1)$-th lattice. Explicitly,
%$(Q f)(l'_\m) = \sum_{l_\n} Q(l'_\m,l_\n) f(l_\n)
%= 2^{-4} \sum_{r_\m=0,1} f(2l'_\m+r_\m)$.
$(Q f)(l') = \sum_{l_\n} Q(l',l) f(l)
= 2^{-4} \sum_{r_\m=0,1} f(2l'+r)$.
This definition implies $c = 2^{-4}$ in Eq.~(\ref{Q}).
Using eq.~(\ref{Dninv}) we obtain
\begin{subequations}
\label{solve}
\begin{eqnarray}
  D_n^{-1}(p) &=& \a_n^{-1}
  -\sum_\m \big( i(\g_\m \otimes I) \ca_\m^n(p)
  + (\g_5 \otimes \t_5 \t_\m) \cb_\m^n(p) \big) \,,
\label{slvD}
\NXT
  \ca_\m^n(p) &=& \sum_{\kn_\m} \ca_\m^0(p+2\p \kn;2^{-n}) |Q_n(p,\kn)|^2\,,
\label{slvA}
\NXT
  \cb_\m^n(p) &=& \sum_{\kn_\m} \cb_\m^0(p+2\p \kn;2^{-n}) |Q_n(p,\kn)|^2\,,
\label{slvB}
\NXT
  |Q_n(p,\kn)|^2 &=& \prod_\n
  \left( \frac{\sin(p_\n/2)}{2^n \sin((p_\n+2\p \kn_\n)/2^{n+1})} \right)^2 \,,
\label{Qn}
\end{eqnarray}
\end{subequations}
where $-\p \le p_\m \le \p$
and for each $\m$, $\kn_\m = -2^{n-1}, -2^{n-1}+1,\ldots,2^{n-1}-1$.
To arrive at Eq.~(\ref{Qn}), observe that we have set $a=1$
for the coarse-lattice spacing,
so that a single-step blocking kernel is a mapping into
this lattice from a lattice with spacing equal to $1/2$.
For each $\m$, this kernel has the momentum representation
$(1/2)(\exp(ip_\m/2) + 1) = \exp(ip_\m/4) \cos(p_\m/4)
= \exp(ip_\m/4) \sin(p_\m/2) / (2 \sin(p_\m/4))$. Eq.~(\ref{Qn}) follows by going
``backwards'' down to the fine lattice with spacing $a_0 = 2^{-n}$.

The massless staggered operator satisfies $\{D_0,(\g_5\otimes\t_5)\}=0$,
and therefore the staggered action is invariant under a $U(1)$
chiral symmetry.\footnote{
  This symmetry correspond to the $U(1)_\e$ symmetry of the one-component
  formalism \cite{kwsm}. Note that the interpretation of this symmetry
  in the continuum limit -- axial, vector or some combination of them --
  depends in general on the choice of the staggered mass term \cite{mgjs}.
  For the simple mass term of Eq.~(\ref{Dst}) it is a chiral symmetry.
}
For the RG-blocked operator we have Eq.~(\ref{slvD}),
which implies that $ D_n^{-1}(x,y)$ anti-commutes with $(\g_5\otimes\t_5)$
except for $x=y$. In fact,
\begin{equation}
  \{D_n,(\g_5\otimes\t_5)\} = 2\a_n^{-1} D_n\,(\g_5\otimes\t_5) D_n \,,
  \qquad n\ge 1 \,.
\label{gwl}
\end{equation}
This is recognized as a generalization of the GW relation \cite{gw}.
Thus, after the first blocking transformation the U(1) chiral symmetry
gets modified to a Ginsparg-Wilson-L\"uscher (GWL) chiral symmetry \cite{l}.

For Wilson fermions, it was proved in ref.~\cite{rgb} that the RG-blocked
operator
\begin{equation}
  D_n(x-y) = \int_{BZ} dp\;
  e^{ip(x-y)} D_n(p)\,,
\label{Dxy}
\end{equation}
is local.
Here $\int_{BZ} dp \equiv \int_{-\p}^{\p} \frac{d^4p}{(2\p)^4}$
denotes the integration over the Brillouin zone
of the coarse lattice. As explained earlier, this means that
$D_n(x-y)$ decays exponentially with $|x-y|$,
and the decay rate is $O(1)$ in units of the coarse-lattice spacing.
The bounds established in
the course of the proof are uniform in $n$, and hold for $\a \le \hat\a$
where $\hat\a>0$ is an $O(1)$ constant whose actual value can be worked out
by keeping track of the details of the proof.

We now argue that the proof of locality continues to hold if,
at the starting point, we replace the Wilson operator
by the staggered Dirac operator~(\ref{Dst}). This amounts
to replacing the Wilson term $W$
by a ``skewed'' Wilson term $W_{\rm st}$, where
\begin{eqnarray}
  W &=& \sum_\m (1-\cos(p_\m a))\,,
\NON
  W_{\rm st} &=& \sum_\m (\g_5 \otimes \t_5 \t_\m) (1-\cos(p_\m a))\,.
\label{WW}
\end{eqnarray}
The proof requires lower
and upper bounds on $W$ as a function of $p_\m a$.
Introducing the vector-space norms
$|x_\m|_\g\, = \big( \sum_\m |x_\m|^\g \big)^{1/\g}$
we observe that $W=|W|=|1-\cos(p_\m a)|_1$.
In the staggered case, we have the operator norm
$\abss{W_{\rm st}}^2\, = W_{\rm st} W_{\rm st}^\dagger
= \sum_\m  (1-\cos(p_\m a))^2$,
or equivalently $\abss{W_{\rm st}}\, = |1-\cos(p_\m a)|_2$.
Since the following equivalence-of-norms inequalities hold in $d$ dimensions
\begin{equation}
  d^{-1/2} |x_\m|_2 \le |x_\m|_1 \le d^{1/2} |x_\m|_2 \,,
\label{norm}
\end{equation}
it follows that every lower or upper bound on $|W|$ entails a corresponding
bound on $\abss{W_{\rm st}}$, and {\it vice versa}.
This simple argument shows that, indeed, the proof given in ref.~\cite{rgb}
generalizes to the RG-blocked staggered Dirac operator
in its flavor representation.

The key physical input that goes into the proof \cite{rgb} is that
$D_0^{-1}(p)$ and $D_n^{-1}(p)$ share the same singularity as $p \to 0$,
namely, $-i\sum_\m \g_\m p_\m/p^2$. Indeed, the singularity of $D_n^{-1}(p)$
arises only from the $\kn_\m=0$ term on the right-hand side of Eq.~(\ref{slvA}).
Factoring out this singularity by writing $D_0(p) = R_n(p) D_n(p)$,
one can prove that the operator $R_n(p)$ is analytic in $p_\m$
and that both $R_n(p)$ and $R_n^{-1}(p)$ are bounded.
This is then used to prove that $D_n(p)$ and $G_n(p)$ are analytic,
and that  $D_n(p)$, $G_n(p)$ and $G_n^{-1}(p)$ are bounded.
Exponential localization of the corresponding coordinate-space
kernels follows from general theorems.

We will next show that, in the limit of infinitely many RG steps,
$D_n$ becomes diagonal in flavor space.
The flavor-mixing part of $D_n^{-1}(p)$ is given by Eq.~(\ref{slvB}),
where explicitly
\begin{equation}
  \cb_\m^0(p+2\p \kn;2^{-n})
  = \frac{2 \sin^2((p_\m+2\p \kn_\m)/2^{n+1})}
  {2^n  \sum_\n \big( \sin^2((p_\n+2\p \kn_\n)/2^n)
  + 4 \sin^4((p_\n+2\p \kn_\n)/2^{n+1}) \big)}\,.
\label{B0}
\end{equation}
It is easy to see that
$|\cb_\m^0(p+2\p \kn;2^{-n})| \le c_1 2^{-n}$ where $c_1=O(1)$.
Away from the singularity at $p+2\p \kn=0$ this is evident.
For $|p+2\p \kn| \ll 2^n$, the same result
follows using $\sin(x) \sim x$ for $x \ll 1$. Thus,
$|\cb_\m^0(p+2\p \kn;2^{-n})| = O(2^{-n})$, uniformly in $p$ and $\kn$.
In addition, $|Q_n(p,\kn)|^2 \le c_2 \prod_\n |p_\n|^2/|(p_\n+2\p \kn_\n|^2$
where again $c_2=O(1)$ \cite{rgb}. Hence the $\kn_\m$-summation converges,
and $|\cb_\m^n(p)| = O(2^{-n})$ for all $p$. Taking the limit $n\to\infty$
we conclude that $\cb_\m^n(p)\to 0$, uniformly in $p$.

The inverse fixed-point operator obtained in the limit $n\to\infty$
can be expressed as
\begin{equation}
  D^{-1}_\infty(p) = D^{-1}_{\rm rg}(p) \otimes I
  = \left(\a_\infty^{-1} -i\sum_\m \g_\m \ca^\infty_\m(p)\right) \otimes I \,,
\label{lim}
\end{equation}
where $\a_\infty = (15/16)\a$, and
$\ca^\infty_\m(p) \sim p_\m /p^2$ for $p_\m \ll 1$.
This shows that $D_\infty$ is diagonal in flavor space
and satisfies the (standard) GW relation.\footnote{
  The relevance of the GW relation for establishing consistency
  of the fourth-root trick was recently pointed out in ref.~\cite{mp}.
}
Letting $G_\infty = \lim_{n\to\infty} G_n$, Eqs.~(\ref{RG}) and~(\ref{lim})
imply
\begin{equation}
  \lim_{n\to\infty} {\rm det}^{1/4}(D_0(a_0=2^{-n}))
  = {\rm det}(D_{\rm rg})\, {\rm det}^{1/4}(G_\infty^{-1}) \,.
\label{fctr}
\end{equation}
In view of the locality and boundedness properties established above,
the desired decomposition~(\ref{decomp}) is achieved in this limit!

The essence of the RG blocking is that it distills the long-distance dynamics,
extracting it out of the underlying short-distance theory.
The long-distance dynamics is contained in $D_{\rm rg}$ which is manifestly
diagonal in flavor space, while all the flavor-mixing effects are contained
in $G_\infty^{-1}$. Since $G_n^{-1}$ is analytic in momentum space and has
an $O(1)$ gap, its fourth root shares similar properties.
Hence $G_n^{-1/4}$ is local, and has only cutoff-mass excitations,
uniformly in $n$.

Let us elaborate on this last statement. Observing that $\a$
has mass dimension equal to one, and focusing \eg on the first blocking step,
an RG transformation works by first replacing $D_0$
with $G_1^{-1}=D_0 +\a Q^\dagger Q$ ({\it cf.} Eq.~(\ref{RGb})).
Now, the massless operator $D_0$ has vanishingly
small eigenvalues near $p_\m=0$. The contribution from the blocking kernel,
$\a Q^\dagger Q$, lifts these small eigenvalues and generates an
$O(\a)=O(1)$ gap in the spectrum of $G_1^{-1}$.
We may define the fourth-root operator for any finite $n$ via
\begin{equation}
  \cm_n(\tx,\ty) = \int_{BZ}^{(n)} d\tp \, e^{i\tp(\tx-\ty)} \cm_n(\tp)\,,
  \qquad
  \cm_n(\tp)=(G_n(\tp) G_n^\dagger(\tp))^{-1/8} \,.
\end{equation}
In this equation, $\tx,\ty$ take values on the fine lattice,
and $\int_{BZ}^{(n)} d\tp$
denotes the integration over the fine-lattice Brillouin zone.
The argument why $\cm_n(\tx,\ty)$ is local is standard \cite{rgb}.
If we let one of the momentum components become complex,
the singularity closest to the real axis will be at a distance
which is $O(\a_n)$. Deforming the contour of integration,
this implies that $\cm_n(\tx,\ty)$ decays exponentially with $|\tx-\ty|$,
with a decay rate which is $O(\a_n)$, namely, $O(1)$ in units
of the {\it coarse}-lattice spacing.

It is interesting to compare this result to ref.~\cite{adams}
which attempts to find an operator $\cn$
such that ${\rm det}(\cn)={\rm det}^{1/4}(D_0)$, without the help
of RG transformations (see also ref.~\cite{sqst}).
In this case, the gap is provided by the {\it physical} mass,
and the decay rate of the square-root kernel needed in the construction
is found to be $O(\sqrt{m/a})$.
Hence, the limit $m\to 0$ is problematic.

With the help of RG blocking, the small-distance scale relevant for the
fourth root is $a_0$, the spacing of the original fine lattice,
and the relevant large-distance scale is $a$,
the spacing of the coarse lattice.
In comparison, in refs.~\cite{sqst,adams} the relevant short-
and long-distances scales are $a$ and $1/m$ respectively. The use of
RG blocking effectively achieves the replacements $a\to a_0$, $1/m \to a$.
While the mathematics is similar, the physical conclusion is different.
The ``fourth-root'' kernel $\cm_n(\tx,\ty)$ is long-ranged with respect
to the fine-lattice scale $a_0$ (in analogy with refs.~\cite{sqst,adams});
but the same kernel is \textit{short-ranged}
with respect to the coarse-lattice scale.
This is true uniformly in $n$, hence also in the limit $n\to\infty$.

The RG analysis generalizes to the case that the
mass term is switched back on in Eq.~(\ref{Dst}).\footnote{
  It is desirable (though \textit{not} a necessary condition for $ma \ll 1$)
  to choose the same relative sign for $m$ and $\a$, such that
  $m+\a Q^\dagger Q$ is a strictly positive operator.
}
We find that Eq.~(\ref{slvD}) still holds, except that the explicit
expressions for $\ca_\m^n(p)$, $\cb_\m^n(p)$ and $\a_n^{-1}$ get modified
(now $\a_n^{-1}= \a_n^{-1}(p)$ becomes a non-trivial function of $p$).
The proof of locality generalizes to $m\ne 0$ for Wilson fermions \cite{rgb},
and the same should be true for the flavor-representation
staggered fermions. Also, clearly
$\cb_\m^0(p;m\ne 0) \le \cb_\m^0(p;m=0)$. Therefore, for $n\to\infty$,
$\cb_\m^n(p)$ tends to zero as before. The limiting operator $D_{\rm rg}$,
defined by the first equality in Eq.~(\ref{lim}),
is again diagonal in flavor space. Of course, for $m\ne 0$, $D_{\rm rg}$
will not satisfy the GW relation any more.

%%%%%%%%%%%%
\vspace{5ex}
\noindent {\large\bf 4.~Interacting staggered fermions}
%\secteq{4}
\vspace{3ex}

It is unlikely that rigorous theorems such as those of ref.~\cite{rgb} will ever
be generalized to an interacting lattice theory.
Still, physical intuition suggests that similar
statements on locality and boundedness may hold true in an
interacting theory as well. In this section, I address some of the issues
that arise when dealing with an interacting theory.

The question is what are the properties of an RG-blocked lattice theory,
when the initial interacting theory involves one-component
staggered fermions, and the fourth-root trick is applied.
In an interacting theory RG transformations
may be realized in numerous ways. A common feature
is that RG transformations naturally give rise to a ``two-cutoff'' theory:
the RG-blocked theory living on a coarse lattice with spacing $a$ is obtained
after applying $n$ RG transformations to an initial theory defined
on a fine lattice with spacing $a_0 \ll a$.
For example, in the context of ``perfect action'' one applies
RG transformations to fermion and gauge fields alike,
and the limit $n\to \infty$ (and $a_0\to 0$) is taken while keeping $a$ fixed
\cite{h}.

Here I will limit the discussion to a simpler framework,
where RG transformations are applied only to the fermion variables.
Among other things, this has the advantage that some simple tests can
be carried out on existing dynamical configurations.

The first problem that must be tackled is that, as discussed below,
the interacting theory is defined using one-component staggered fermions
for a reason \cite{mgjs,mw}.
In the free-field case, there is a unitary operator $Q_0$
that maps the one-component staggered operator, defined on a lattice
with spacing $a_0$, to the flavor-representation staggered operator
on a lattice with spacing $a_1=2a_0$ \cite{fr}.
In the interacting case, the mapping must preserve gauge invariance
but there is no unique, obvious way to define it.

We propose to deal with this problem by a single RG blocking step
which keeps the number of fermionic degrees of freedom unchanged.
Specifically, in Eq.~(\ref{RG}) we take $\j,\bj$ to be single-component fields,
while $\c,\bc$ are four flavors (or ``tastes'') of Dirac fields.
$D_0$ is now a covariant, interacting one-component staggered operator.
For the blocking kernel we choose some covariant version of $Q_0$,
denoted $\cq^{(0)}$ below, defined in terms of the link
variables on the fine lattice.
Eqs.~(\ref{RG}) and~(\ref{DG1}) provide explicit expressions for $D_1$,
which is the resulting staggered operator in the flavor representation,
as well as for $G_1$,  whose (inverse) determinant is picked up
when performing this non-trivial change of variables.
(We will shortly return to the role of ${\rm det}(G_1^{-1})$.)
Using the natural embedding $x\to\tx$ from the coarse lattice
to the fine lattice (as described above Eq.~(\ref{solve})),
both $D_1$ and $G_1$ are gauge-covariant
functions of the link variables on the fine lattice.

Having thus constructed a ``flavor representation'' in the interacting theory,
we may apply $n$ additional,
ordinary RG blocking transformations to the fermions.
These block transformation dilute the number of fermionic degrees of freedom
while maintaining gauge invariance, provided that we keep choosing
blocking kernels which are covariant with respect to
gauge transformations on the original fine lattice.
This is naturally realized if, for any
point $x'$ and any point $y$ such that $Q^{(j)}(x',y) \ne 0$ in the
free theory, we construct the covariant kernel  $\cq^{(j)}(\tx',\ty)$
by summing over Wilson lines that go
from $\tx'$ to $\ty$ on the original fine lattice.

An important question is what is the fate of the global symmetries of
the original, interacting one-component staggered-fermion theory.
As for the chiral $U(1)$ symmetry, we have seen
that it becomes a GWL symmetry (Eq.~(\ref{gwl})).
Now, it may be impossible to preserve manifest hypercubic invariance in the
construction of $\cq^{(j)}$, and the same is true for the
staggered shift symmetry \cite{mgjs}.
Since the original one-component theory has exact hypercubic and shift
symmetries, by Eq.~(\ref{RGd}), any breaking of these symmetries
induced in the above-defined flavor representation by $D_n$
must be exactly compensated by ${\rm det}(G_n^{-1})$.
In other words, the effective action
$S_{\rm eff}^n=\log({\rm det}(G_n))$ should automatically contain the
local ``counter-terms'' needed to restore exact invariance.

It is known that the attempt to construct an interacting theory directly in the
flavor representation gives rise to induced cutoff-scale masses
which, on top of that, violate Lorentz invariance \cite{mw}.
The interacting staggered theory is defined in the one-component formalism
because its global symmetries, including in particular
shift symmetry, forbid these disastrous mass terms \cite{mgjs}.
It is therefore crucial that the cancellation mechanism proposed above
will indeed be operative when the flavor representation
is constructed though ``RG blocking'' from the one-component theory.
Interestingly, there exists a general RG-blocking result which provides
direct evidence that the above cancellation mechanism
indeed works as it should. Introduce the ``telescopic sum'' \cite{rgb}
\begin{eqnarray}
  D_0^{-1} &=& D_0^{-1} Q_n^\dagger D_n Q_n D_0^{-1}
  + \sum_{j=0}^{n-1} D_0^{-1} \big(
  Q_j^\dagger D_j Q_j - Q_{j+1}^\dagger D_{j+1} Q_{j+1}\big) D_0^{-1}
\NON
  &=& S_n D_n^{-1} \bs_n + \sum_{j=0}^{n-1} S_j \tG_{j+1} \bs_j \,,
\label{tele}
\end{eqnarray}
where we have set $Q_0=I$ and where
\begin{equation}
  S_j \equiv D_0^{-1} Q_j^\dagger D_j\,, \qquad
  \bs_j \equiv D_j Q_j D_0^{-1}\,,
\label{Sj}
\end{equation}
and
\begin{equation}
  \tG_{j+1}
  \equiv D_j^{-1} - D_j^{-1} Q^{(j+1)\dagger} D_{j+1} Q^{(j+1)} D_j^{-1}
  = \G_{j+1} \,.
\label{Gammaj}
\end{equation}
The equality $\tG_{j+1}=\G_{j+1}$ follows by substituting Eq.~(\ref{Djinv})
for $D_{j+1}$ and re-expanding as a geometric series.

Consider now an interacting staggered-fermion theory
with light or massless quarks. Let us examine
the long-distance behavior of the various kernels in Eq.~(\ref{tele}).
In ref.~\cite{rgb} it is proved that, in the free theory,
$S_j$, $\bs_j$ and $\G_j$ all have ranges which are $O(1)$
in units of the coarse-lattice spacing. We expect the same
to hold in the interacting case.
Because the original kernel $D_0^{-1}(\tx,\ty)$
is long-ranged in {\it physical} units,
the only way for Eq.~(\ref{tele}) to hold is if
$D_n^{-1}(\tx,\ty)$ is long-ranged too.

The physical interpretation of Eq.~(\ref{tele}) is not completely straightforward
because $D_0^{-1}$ and $D_n^{-1}$ are not gauge invariant all by themselves.
Let us examine a physical observable.
We will consider the gauge-invariant two-point function
of the true Goldstone-boson field $\p(x)$.
Its existence is implied by the symmetries of the
one-component formalism \cite{mgjs}, and so
$\svev{\p(x)\,\p(y)}$ decays like a power of $|x-y|$ in the massless limit.
We may construct this two-point function
either from $D_0^{-1}$ or, using Eq.~(\ref{tele}), from $D_n^{-1}$.
Had any cutoff-scale masses been induced as in the case of ref.~\cite{mw},
$D_n^{-1}$ would decay exponentially with a cutoff-scale rate,
making it impossible to reproduce the correct long-distance behavior.
We conclude that, whether or not the RG-blocked theory
enjoys manifest shift and (full) hypercubic symmetries,
the physical consequences of all the original symmetries remain intact!

Suppose that, instead, one had started with an interacting theory obtained
by simply gauging the flavor representation
using link variables that reside on the coarse lattice, as in ref.~\cite{mw}.
Assume also that the bare quark mass is set to zero.
What would go wrong? Denoting the new interacting Dirac operator by $\cd_0$
and the lattice spacing by $a_0$, The two-point function of the
would-be Goldstone boson is given explicitly by
\begin{subequations}
\label{wouldbe}
\begin{eqnarray}
  \svev{\p(x)\,\p(y)} &=& \svev{\cg(x,y)} \,,
\label{pipi}
\NXT
  \cg(x,y) &=& {\rm tr} \left((\g_5 \otimes \t_5) \cd_0^{-1}(x,y)
                              (\g_5 \otimes \t_5) \cd_0^{-1}(y,x)\right) \,.
\label{Gpipi}
\end{eqnarray}
\end{subequations}
Since $(\g_5 \otimes \t_5) \cd_0 (\g_5 \otimes \t_5) =\cd_0^\dagger$
one has $\cg(x,y)>0$, \ie this correlator is strictly positive.
The decay rate of $\svev{\p(x)\,\p(y)}$ is now $O(1/a_0)$,
because the explicit one-loop calculation of ref.~\cite{mw} shows that the quarks
acquire (Lorentz-breaking) cutoff masses in this theory.
Since $\cg(x,y)>0$, it is ruled out that the short range could be generated by
destructive interference between different gauge-field configurations;
rather, given an ensemble of configurations,
we must have $\abss{\cd_0^{-1}(x,y)}^2\, \sim \exp(|x-y|/a_0)$
on each and every one of them.

One could have applied RG transformations, and
Eq.~(\ref{tele}) would again say that any long-distance physics of $\cd_0$
must be reproduced by the resulting $\cd_n$.
However, we have just seen that $\cd_0$ has
{\it no} long-distance physics whatsoever.
By Eq.~(\ref{tele}), any number of RG transformations would not have
fixed it.

Coming back to the interacting one-component staggered theory,
the algebraic structure of $\cg(x,y)$ is similar to Eq.~(\ref{Gpipi}),
and again $\cg(x,y)>0$. In order to obtain the correct power-law
behavior of the Goldstone-boson correlator $\svev{\p(x)\,\p(y)}$,
there must exist a finite probability
to encounter configurations where $\cg(x,y)$ has the same long-distance
behavior. Finally, by Eq.~(\ref{tele}), the long-distance behavior
must be sustained if we use $D_n^{-1}$ instead of $D_0^{-1}$
to construct $\cg(x,y)$. The long-distance physics contained in $D_0^{-1}$
is faithfully reproduced by the RG-blocked $D_n^{-1}$.

It will be interesting to test the resulting operators numerically,
and see  whether the properties established in the free theory persist.
For the RG-blocked operator $D_n$,
the crucial properties are locality, suppression of the flavor-mixing
(or taste-mixing) part with $n$, and convergence to a GW operator
in the massless case.
As for $G_n$ (Eq.~(\ref{Gn})), the question is whether this operator indeed
describes only excitations with masses of the order of the cutoff,
so that the effective action $S_{\rm eff}^n=\log({\rm det}(G_n))$ is local.
Last, it should be verified that the mechanism that protects
the (physical consequences of the) symmetries of the original one-component
formulation is indeed operative.

On the theoretical side, an interesting  idea is to make use of the notion of
\textit{admissibility condition}. The original concept introduced in
ref.~\cite{hjl} asserts that the lattice gauge field is constrained such
that, for every plaquette, ${\rm Re\, tr}(1-U_{\m\n}(x)) < \e_0$,
where $U_{\m\n}(x)$ is the ordered product of link variables
around the given plaquette, and $\e_0>0$ is a fixed (small) number.
It is believed that an admissibility condition does not change
the universality class.
The utility of an admissibility condition is the following.
Given a free lattice operator whose spectrum satisfies a certain bound,
one expects that  this bound will be modified only by $O(\e_0)$
if we promote the operator to a covariant one, while
allowing only for gauge fields that satisfy the admissibility condition.

In the case at hand, we may envisage imposing an admissibility condition
on the link variables of the original fine lattice. This should imply
that, apart from $O(\e_0)$ modifications,
the operator bounds established in ref.~\cite{rgb} will continue to hold
during the first few blocking steps, and the same should follow for
the locality properties.

Unlike in ref.~\cite{hjl}, however, we now face the following problem.
When the number of blocking transformations
becomes of the order of  $n\sim 1/\e_0$, we no longer have any useful bound
on the gauge field. We will thus propose a stronger notion
of admissibility. Considering the product of (covariant) blocking kernels
$\cq_n=\cq^{(n)} \cq^{(n-1)}\cdots \cq^{(1)} \cq^{(0)}$,
we will constrain the gauge field on the original fine lattice
by demanding that every Wilson loop
$\cw$ occurring in the product $\cq_n \cq_n^\dagger$ will satisfy
the constraint $|\cw|< \e_0$.
The reasoning behind this new admissibility condition is the following.
After $n$ blocking steps, the length of the loops contained in
$\cq_n \cq_n^\dagger$ will be $O(2^n)$ in units of the original lattice spacing
$a_0$. However, as explained earlier, we are really interested in the
``two cutoff'' situation where $a_0 = 2^{-n}a$. When measured in units
of the coarse-lattice spacing $a$, all these Wilson loops have length
smaller than some $l_0=O(1)$. Our new admissibility condition is therefore
a natural generalization of the same concept to a ``two cutoff'' situation.
With this new admissibility condition, it is plausible that the free-theory
bounds will continue to hold for arbitrarily large $n$, up to $O(\e_0)$
modifications. In other words, the deviations from the free-theory bounds
will be $O(\e_0)$ independently of $n$.
In addition, for  very large $n$,
the flavor-breaking part of the RG-blocked operator will become very small.
It should be possible to find large enough, but finite, $n$, such that
full flavor (``taste'') symmetry is recovered to any
desired accuracy.\footnote{
  Since in this framework the independent link variables always reside on the
  original fine lattice, the limit $n\to\infty$ cannot be taken
  independently of the continuum limit.
}

In conclusion, using the machinery of RG block transformations
I have shown that the fourth-root trick is consistent for free staggered
fermions. In the limit of infinitely many RG steps,
the free staggered-fermion determinant is equal to the fourth power
of the determinant of a one-flavor local operator which,
in the massless case, satisfies the GW relation,
times the determinant of a local operator whose excitations have cutoff masses.
The fourth root of the latter operator is local too.
While a similar result for interacting staggered fermions
is unlikely to be established anytime soon, I have discussed
how to construct the flavor representation in the interacting theory,
while in effect maintaining all the symmetries of the one-component
formalism. I have also suggested avenues
for numerical tests, as well as a theoretical framework which appears to
be best suited for generalizing some of the rigorous free-theory results
to the interacting case.

%%%%%%%%%%%%
\vspace{5ex}
\noindent {\bf Acknowledgements}
\vspace{3ex}

I thank Claude Bernard and Maarten Golterman for extensive discussions.
This work was initiated during the workshop
``Matching light quarks to hadrons" at the Benasque Center for Science
in Spain, and I wish to thank both the participants and the organizers
for creating an exciting atmosphere.
This work is supported by the Israel Science Foundation under grant
no. 222/02-1.

%%%%%%%%%%%%

\end{document}